\documentclass[final,english]{bullsrsl}

\usepackage[latin1]{inputenc}

\usepackage[T1]{fontenc}

\usepackage{natbib} 

\usepackage{graphicx}

\begin{document}
\title{The influence of free-free absorption on the radio spectrum of Particle-Accelerating Colliding-Wind Binaries}



  \author[affil={1}, corresponding]{Micha\"el}{De Becker}
  \author[affil={1}]{Agustina Belen}{Blanco}
  \author[affil={1}]{Mathilde}{Tasseroul}
  \author[affil={2}]{Anandmayee}{Tej}
  \author[affil={2,3}]{Anindya}{Saha}
  \author[affil={4}]{Paula}{Benaglia}
  \author[affil={5}]{Santiago}{del Palacio}

\affiliation[1]{Space sciences, Technologies and Astrophysics Research (STAR) Institute, University of Li\`ege, Belgium.}
\affiliation[2]{Indian Institute for Space science and Technology (IIST), Thiruvananthapuram, India}
\affiliation[3]{Instituto Argentino de Radioastronom\'{\i}a, Villa Elisa, Buenos Aires, Argentina}
\affiliation[4]{Kavli Institute for Astronomy and Astrophysics, Peking University, People's Republic of China}
\affiliation[5]{Department of Space, Earth and Environment, Chalmers University of Technology, Gothenburg, Sweden}


\correspondance{Michael.DeBecker@uliege.be}


\maketitle

\begin{abstract}
The study of massive stars in binary and higher-multiplicity systems that participate in particle acceleration is a key topic at the crossroads of massive star physics, shock physics, and galactic cosmic ray astrophysics. From an observational perspective, radio measurements are our primary tool for identifying these systems through their synchrotron radio emission. Out of the 54 such systems known to date, all but two have been discovered via their non-thermal radio emission. However, identifying these systems is not straightforward. The main challenge lies in free-free absorption, which can obscure the synchrotron signature and hinder detection. This paper summarizes recent developments in our understanding of these systems, with a particular focus on the strong observational bias introduced by free-free absorption. This bias significantly limits our ability to determine the true fraction of particle accelerators among colliding-wind binaries.
\end{abstract}





\section{Scientific context}
Massive stars (O-type, early B-type, and their evolved counterparts such as Wolf-Rayet (WR) stars) are continuously driving strong stellar winds with terminal velocities of 2000--3000 km/s and mass-loss rates in the range of $10^{-7}$--$10^{-5}$ M$_\odot$/yr \citep{Muijres2012,Hamann2019,Sander2019}. The mechanical energy carried by these winds per unit time, i.e., the wind kinetic power, constitutes a valuable energy reservoir available to feed processes of high relevance in massive star physics. A very important feature among massive stars is their multiplicity: a substantial fraction of these stars are found in binary and higher multiplicity systems \citep{Offner2023}. In multiple systems, their winds collide and produce strong shocks. Thus, these systems offer perfect conditions for investigating various aspects of shock physics. Among these processes of interest, Diffusive Shock Acceleration \citep{Drury1983} is likely responsible for the acceleration of particles up to relativistic velocities, as evidenced in 54 systems, hence the Particle-Accelerating Colliding-Wind Binary status (Catalogue of PACWBs, \citealt{DeBecker2013}; \url{www.astro.uliege.be/~debecker/pacwb}).

Most particle accelerators among massive binaries are identified by the detection of synchrotron radio emission produced by relativistic electrons accelerated by shocks associated with colliding winds \citep{Abbott1984,Williams1997,Benaglia2006,Benaglia2020,Montes2009,Marcote2021,DeBecker2024,Benaglia2025}, and this has been the topic of several important theoretical studies over the past decades \citep{Doug,Pit2,Pittard2021}. Investigating colliding-wind binaries in the radio domain thus constitutes the main approach to identify their particle accelerator status \citep{DeBecker2013,DeBecker2017}. However, in contrast to most regular synchrotron emitters, PACWBs display a more complex behavior. Not only are they composite radio emitters displaying both thermal and non-thermal emission, but the detectability of their synchrotron radiation is highly affected by free-free absorption (FFA). 
The latter aspect is the focus of this paper, which aims to provide a summary of the main trends likely to appear when investigating colliding-wind binaries in the radio domain, with an emphasis on the consequences for their identification as particle accelerators.
The strong influence of FFA in this framework is at the root of a strong bias in determining the fraction of PACWBs among massive binary systems \citep{DeBecker2017}. A better view of how FFA affects their radio spectrum is relevant both for (i) understanding how it leads us to underestimate the actual population of PACWBs (relevant for the question of their contribution to the production of Galactic cosmic rays, \citealt{DeBecker2024BINA3}) and for (ii) interpreting their radio spectra as a whole, as addressed in Sect.\,\ref{InflFFA}.

\section{Influence of free-free absorption}\label{InflFFA}
\begin{figure}
\centering
\includegraphics[width=0.9\textwidth]{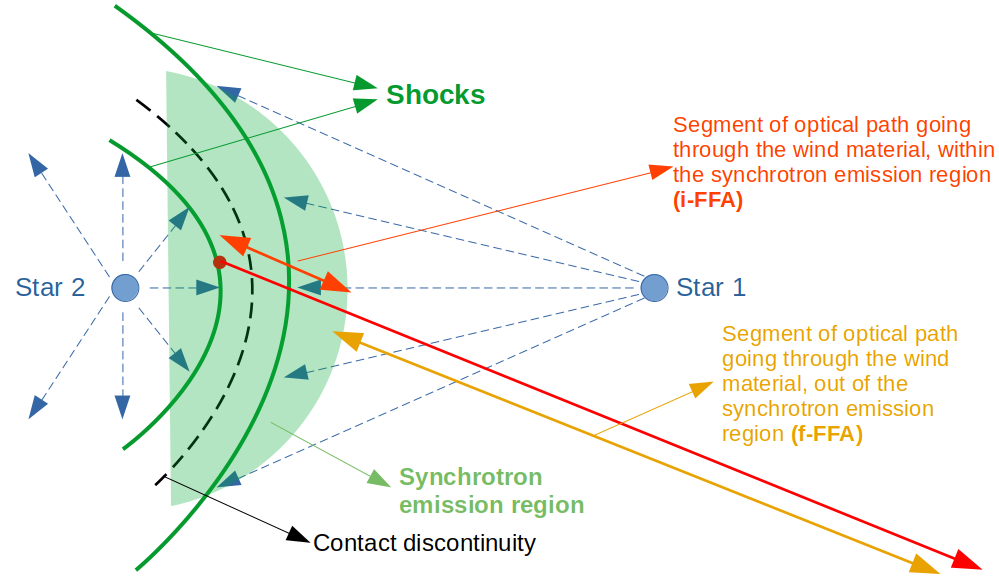}
\begin{minipage}{12cm}
\centering
\caption{Illustration of a PACWB where the wind of Star 1 dominates that of Star 2, leading to a bow-shape interaction region partly folded about the position of Star 2. The dashed blue arrows represent some wind flow lines heading to the shocks. The red arrow illustrates a specific optical path of synchrotron radiation produced at the position marked with the red dot. The fraction of optical path outside the synchrotron emission (green area) region is responsible for f-FFA, while the wind material spatially coincident with the NT emission region leads to i-FFA.}\label{FFA-i-f}
\end{minipage}
\end{figure}

\begin{figure}
\centering
\includegraphics[width=0.7\textwidth]{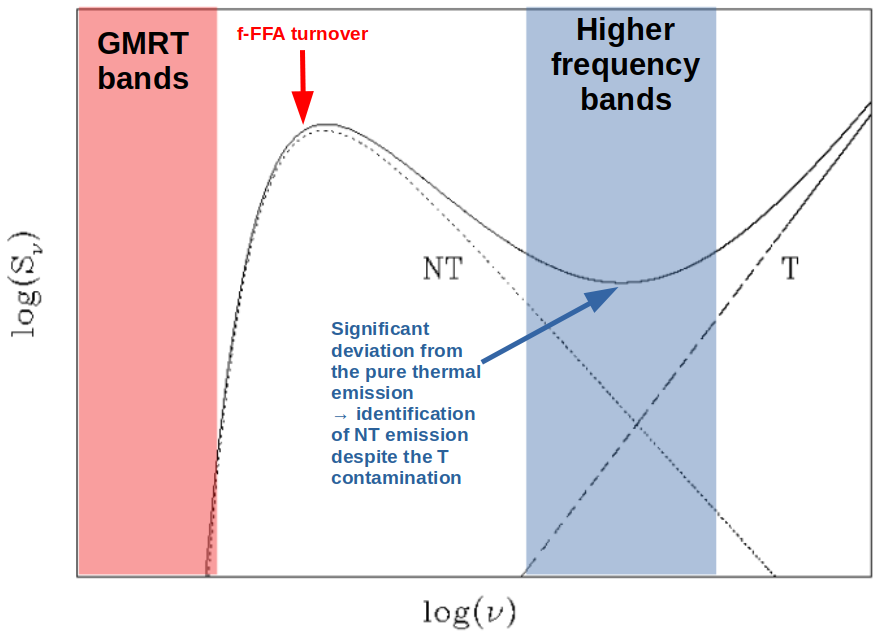}
\begin{minipage}{12cm}
\centering
\caption{Generic radio spectral energy distribution of a PACWB where the low-frequency part is severely attenuated by foreground FFA. For illustration purposes, red and blue vertical strips illustrate the bandpass covered by the GMRT and other radio facilities at GHz frequencies, respectively. The red arrow points to the position of the turnover frequency that is slightly lower than that of the peak of the spectrum. Figure adapted from \citet{Blanco2024}.}\label{SEDfFFA}
\end{minipage}
\end{figure}

The basic geometry of a PACWB is illustrated in Fig.\,\ref{FFA-i-f}, where the effect of FFA is considered along a specific optical path going through the interaction region and the surrounding wind material. The strong asymmetry typically illustrates the case of a WR + O system, where the WR (Star 1) wind dominates that of the O-star (Star 2). The strong FFA by the wind material (especially by a WR star) along the line-of-sight can severely attenuate synchrotron radiation produced around the colliding-wind region. This typically happens when the foreground plasma that the stellar winds are made of is significantly dense. In that case, one refers to foreground FFA (f-FFA). The generic spectral energy distribution of a PACWB is shown in Fig.\,\ref{SEDfFFA}. Individually, and in the absence of any wind collision, stellar winds produce free-free emission distributed as a power-law with a positive index \citep{WB,PF}. In contrast, the synchrotron emission component presents a negative slope in the $\log-\log$ space, such that the non-thermal flux density scales as $S_\nu \propto \nu^{-\alpha}$ (where $\nu$ is the photon frequency and $\alpha \sim 0.5$-1 is the spectral index). Synchrotron radiation is therefore expected to dominate at lower frequencies. However, it is also at low frequencies that f-FFA leads to an exponential cut-off capable of suppressing the synchrotron component from the spectrum. This turnover process is very likely the main one that prevents the detection of synchrotron radio emission in massive binaries, as claimed, for instance, in studies aimed at searching for the signature of particle acceleration using the upgraded Giant Meterwave Radio Telescope (uGMRT) \citep{Saha2023,Blanco2024}, or even at frequencies of several GHz where FFA is expected to be less severe \citep{DeBecker2019}. 

\begin{figure}
\centering
\includegraphics[width=0.7\textwidth]{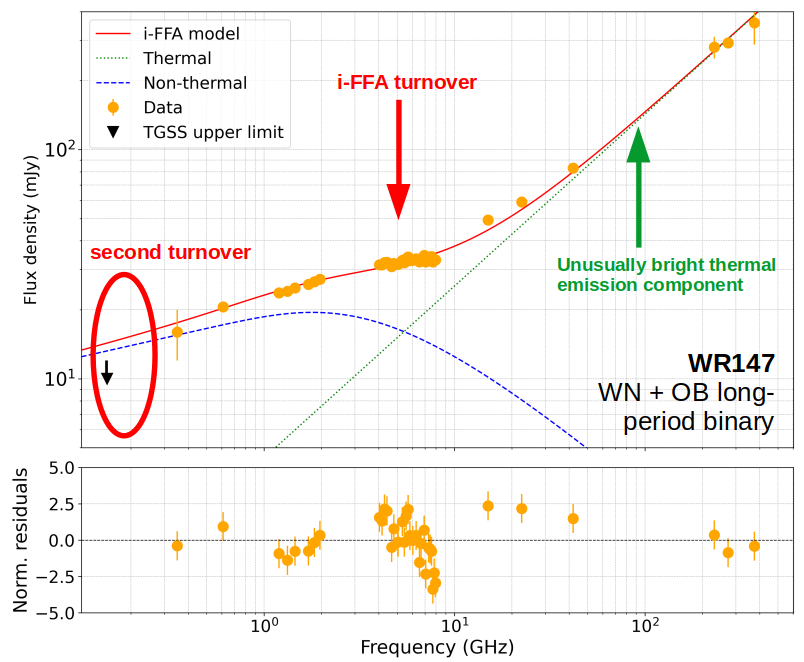}
\begin{minipage}{12cm}
\centering
\caption{Upper panel: radio spectral energy distribution of WR~147 fitted by a composite internally free-free absorbed model. In the legend, TGSS stands for TIFR GMRT Sky Survey. The red arrow is located at the position of the turnover frequency for i-FFA as obtained by \citet{Tasseroul2025}. Lower panel: normalized residuals in the sense data minus model. Figure based on the data published by \citet{Tasseroul2025}.}\label{SEDwr147}
\end{minipage}
\end{figure}

In some cases, FFA occurs because the material coincides spatially with the synchrotron emission region, i.e., internal FFA (i-FFA). This leads to a less severe but still significant attenuation below a given turnover frequency. The emblematic example is WR~147 \citep{Tasseroul2025}, and its spectral energy distribution is shown in Fig.\,\ref{SEDwr147}. The internal absorption is only revealed when no strong f-FFA dominates the absorption below the turnover frequency. This does not prevent a lower amplitude f-FFA from occurring at lower frequencies, as suggested by the radio spectral energy distribution of WR~147. However, it is difficult to clarify the nature of the second low-frequency turnover. Although FFA should be responsible for that second turnover, one cannot completely reject the idea that the Razin-Tsytovitch effect \citep{Pacholczyk1970} is responsible for it. The latter process consists of the inhibition of the synchrotron emission mechanism due to the suppression of the relativistic beaming that drives the emission process. It is thus the result of aborted emission and not of absorption subsequent to the emission. This occurs when the synchrotron emission region coincides with a thermal plasma. It leads to a low-frequency exponential drop in the synchrotron spectrum at a frequency that depends on the local thermal electron density and magnetic field strength \citep{Pacholczyk1970}. 

A strong requirement for i-FFA is a stellar separation in the system large enough to prevent f-FFA from dominating the absorption. Indeed, if the internally free-free absorbed emission region is behind a dense absorbing layer, one recovers the typical exponential cut-off displayed in the spectral energy distribution illustrated in Fig.\,\ref{SEDfFFA}. In addition, an orbital plane almost viewed face-on is certainly another favorable circumstance, as an edge-on view would likely lead to more absorbing wind material being distributed all around the stars in the orbital plane.

\section{Concluding remarks}
\begin{enumerate}
\item FFA is by far the main process that hampers the identification of synchrotron radio emission from massive binaries, as it has the potential to severely attenuate or even suppress the synchrotron signature in colliding-wind binaries. Its influence is expected to depend on the orbital phase, especially in eccentric systems, and may even suppress any hint of synchrotron radiation across the full orbit for too short period systems.
\item Depending on the spatial location/expansion of the absorbing material (thermal plasma), different cases can be considered,
\begin{enumerate}
\item[2.a.] f-FFA occurs when the absorbing material is along the line-of-sight, as the synchrotron source is in the background: this leads to an exponential cut-off of the synchrotron spectrum at low-frequencies. This scenario is likely the most frequent among PACWBs.
\item[2.b.] i-FFa is due to the absorbing material being spatially coincident with a significant part of the synchrotron emission region. This can be explained by a physical extension of the population of emitting relativistic electrons overlapping with some pre-shock plasma that is still at a temperature low enough to allow FFA to operate (the post-shock plasma is likely too hot to contribute significantly). In this scenario, the attenuation is less severe than in the case of f-FFA; therefore, it still allows the synchrotron excess in the radio emission to be measured. Some PACWBs can thus be identified as significant synchrotron emitters at low-frequency radio bands despite the influence of FFA.
\end{enumerate}
\item The influence of FFA, in diversified configurations, is a key factor to be taken into account in the interpretation of the radio spectra of massive binaries. As a consequence, this is a crucial aspect of the identification of particle accelerators among colliding-wind binaries.
\end{enumerate}


\begin{acknowledgments}
This research is part of the PANTERA-Stars collaboration, an initiative aimed at fostering research activities on the topic of particle acceleration associated with stellar sources: \url{www.astro.uliege.be/~debecker/pantera}.  This publication benefits from the support of the Wallonia-Brussels Federation (Belgium) in the context of the FRIA Doctoral Grants awarded to ABB and MT. This research has made use of NASA's Astrophysics Data System Bibliographic Services. This work is a result of the Belgo-Indian Network for Astronomy and Astrophysics (BINA), which promotes astronomical collaborations between Indian and Belgian partners and received financial support from the International Division of the Department of Science and Technology (DST, Government of India) and the Belgian Federal Science Policy Office (BELSPO, Government of Belgium) during the period 2016-2023 through different bilateral projects.
\end{acknowledgments}

\begin{furtherinformation}

\begin{orcids}

  \orcid{0000-0002-1303-6534}{Micha\"el}{De Becker}
  \orcid{0009-0007-5788-1629}{Agustina}{Blanco}
  \orcid{0009-0004-0285-9828}{Mathilde}{Tasseroul}
  \orcid{0000-0001-5917-5751}{Anandmayee}{Tej}
  \orcid{0000-0002-9793-3039}{Anindya}{Saha}
  \orcid{0000-0002-6683-3721}{Paula}{Benaglia}
  \orcid{0000-0002-5761-2417}{Santiago}{del Palacio}
\end{orcids}

\begin{authorcontributions}
This work is part of a long term and collective effort with contributions from all co-authors.
\end{authorcontributions}

\begin{conflictsofinterest}
The authors declare no conflict of interest.
\end{conflictsofinterest}

\end{furtherinformation}



%

\bibliographystyle{bullsrsl-en}

\bibliography{extra.bib}

\end{document}